\documentstyle[aps,prb,epsfig,twocolumn]{revtex}

\def\be{\begin{equation}}
\def\ee{\end{equation}}
\def\eps{\epsilon}
\def\epsm{\eps_{\rm m}}
\def\cl{c_{\rm light}}

\begin{document}
\title{Numerical studies of left-handed materials and arrays of split ring
resonators. }
\author{P. Marko\v{s}$^*$ and C.M. Soukoulis\\
Ames Laboratory and Department of Physics and Astronomy, ISU, Ames, Iowa 50011}

\maketitle

\abstract{We present numerical results on the transmission properties of the left-handed materials (LHM) and split-ring resonators (SRR). The simulation results
are in qualitative agreement with experiments. The dependence of the transmission through LHM on the real and imaginary part of the electric permittivity of the metal, the length of the system, and the size of the unit cell are presented.
We also study the dependence of the resonance frequency of the array of SRR
on the ring thickness, inner diameter, radial and azimuthal gap, as well as on the
electrical permittivity of the board and the embedding medium, where SRR resides.
Qualitatively good agreement with previously published analytical results is obtained.}
\medskip

\noindent{PACS numbers: 73.20.Mf,41.20.Jb,42.70Qs}

\medskip

\section{Introduction}

Very recently, a new area of the research, called left-handed materials (LHM) has been experimentally demonstrated by Smith et al.
\cite{padilla,14}
based on the work of Pendry et al. \cite{6a,ieee}.
LHM are by definition composites, whose properties are not determined
by the fundamental physical properties of their constituents but by
the shape and distribution of specific patterns included in them.
Thus, for certain patterns and distribution, the measured effective
permittivity $\eps_{\rm eff}$ and
the effective permeability $\mu_{\rm eff}$ can
be made to be less than zero.  In such materials, the phase and group
velocity of an electro-magnetic wave propagate in opposite directions
giving rise to a number of novel properties \cite{8}.  This behavior has
been called ``left-handedness'', a term first used by Veselago \cite{9} over
thirty years ago, to describe the fact that the electric field,
magnetic intensity and propagation vector are related by a
left-handed rule.

	By combining a 2D array of split-ring resonators (SRRs) with
a 2D array of wires, Smith et al. \cite{padilla} demonstrated for the first time
the existence of left-handed materials.  Pendry et al. \cite{ieee} has
suggested that an array of SRRs give an effective $\mu_{\rm eff}$, which
can be
negative close to its resonance frequency.  It is also well known \cite{6a,3}
that an array of metallic wires behaves like a high-pass filter,
which means that the effective dielectric constant is negative at low
frequencies.  Recently, Shelby et al. \cite{10} demonstrated
experimentally that the index of refraction $n$ is negative for a LHM.
Negative refraction index was obtained analytically\cite{negn}
and also from numerically  simulated data \cite{we}.
Also, Pendry \cite{11} has suggested that a LHM with negative $n$ can
make a perfect lens.

Specific properties of LHM makes them interesting for 
physical and technological applications. 
While  experimental preparation of the
LHM structures is rather difficult, especially when isotropic structures
are required, numerical simulations could predict how the transmission
properties  depends on various structural parameters of the system.
It will be extremely difficult, if not impossible, to
predict the transmission properties of such materials analytically.
Mutual electro-magnetic  interaction of neighboring SRRs and wires 
makes the problem even more difficult. 
Numerical simulations of various configurations
of SRRs and of LHMs could be therefore very useful in searching of the direction of 
the technological development.

In this paper, we present systematic numerical results for the transmission 
properties of LHMs and SRRs. 
An improved version of the transfer-matrix method (TMM) is used. 
Transfer matrix  was applied to  problems of the transmission
of the electro-magnetic (EM) waves through non-homogeneous  
media many years ago \cite{122,pendry,PW}.  
It was also used in  numerical simulations of the photonic band gap materials
(for references see Ref. 15). 
TMM enables us to
find a transmission  and a reflection matrices from which the transmission,
reflection and  absorption  could be obtained.
The original numerical algorithm was described in Ref. 13.
In  our program 
we use a  different  algorithm which was originally developed for
the  calculation of the electronic conductance
of  disordered solids \cite{12}.  

The paper is organized as follows:
In Section \ref{2} we describe briefly the structure.
We concentrate on the structure displayed in Figure 1.  
In Section \ref{3} we present and discuss our results. 
The dependence of the transmission of the LHM and SRR on the electrical permittivity 
of the metallic components of our structure is given in Section \ref{3a}.
In Section \ref{3b} we present the dependence of the transmission of the LHM on the size of the unit cell and size of the metallic wires. 
In Section \ref{3c} we show the dependence of the resonance frequency of SRR 
on the parameters of the SRR. 
Section \ref{3d} deals with the 
dependence of the resonance frequency on the permittivity of the board 
and embedding media.
In Section \ref{4} we summarize our results and give some conclusions.
Finally, in the Appendix A we give detailed description of the transfer matrix method.

\section{Structure of the LHM meta-material}\label{2}

Both in the experiment and in the numerical simulations, the left-handed 
meta-materials consist
from an array of unit cells, each containing one SRR and one wire. 
Figure 1a shows a realization of the unit cell that we have simulated. 
The size of the unit cell $L_x\times L_y\times L_z$ and the size of SRR itself
are of order of mm.
Waves propagate along the $z$-direction. The SRR lies in the $yz$ plane, 
and the wire is parallel to the $y$ axis. 

As we are interested mostly in the transmission properties of the 
left-handed meta-material,
the configuration as presented in Figure 1a, should be considered 
as one-dimensional. 
Indeed, such  meta-materials  possesses the left-handed properties only 
for the electro-magnetic wave incoming in the $z$-direction and even then only
for a given polarization. 
Two-dimensional structures have been realized in
experiments
\cite{14,10}, in which two SRRs have been positioned in each unit cell
in two perpendicular planes. For such structures, left-handed transmission 
properties have been observed  for waves coming from any direction in the $xz$ plane.
No three - dimensional structure has been realized so far.
%In this paper we concentrate on the numerical analysis of one-dimensional structures.

Figure 1b shows a single square SRR of the type used
for our simulations and also  for experiments \cite{14}. 
The structure of the SRR is defined by the following parameters:
the ring thickness  $c$,
the radial gap  $d$,
the azimuthal gap  $g$ and 
the inner diameter  $r$.
The size of the SRR is
\be\label{ww}
w=4c+2d+r.
\ee

Another parameter is the thickness of the SRR itself (in the $x$-direction).
This thickness is very small in the experiments ($\sim 0.02$ mm).
We can not simulate such thin structures yet. 
In numerical simulations, we divide the unit cell into $N_x\times N_y\times N_z$
mesh points. For homogeneous discretization, used throughout this paper, 
the discretization  defines the minimum unit length  $\delta=L_x/N_x$. All length
parameters are then given as integer of $\delta$. This holds also
for the thickness of the SRR.
Generally, the thickness of SRR used in our
simulations is 0.25-0.33 mm.  Although we do not expect  that the
thickness will  considerably influence the electro-magnetic properties of the SRR,
it still could cause small quantitative difference between our data and
the experimental results.

\section{Structural parameters}\label{3}

\subsection{Metallic permittivity}\label{3a}

The existence of LHM has been experimentally 
demonstrated \cite{padilla,14} for structures that 
have resonance frequencies in the GHz region. 
In this frequency region,
we do not know the exact values of electrical  permittivity $\epsm$ 
of the metal. We know that Im $\epsm$ is very large, and/or the Re $\epsm$
is large but negative. In our previous studies \cite{ms} we have found that the
resonance frequency $\nu_0$ of the LHM depends only on the absolute value of
$\epsm$. In fact $\nu_0$ reaches the saturated value provided that $|\epsm|>10^4$.
Since we do not know the exact values of the metallic permittivity, we have studied the transmission of the LHM with different values of $\epsm$.
In the results presented in Figure 2, we choose
$\epsm=1+i~{\rm Im}~\epsm$ with different values of Im $\epsm$. \cite{kvak} 
The last is proportional to $\sigma(\omega)/\omega$ \cite{jackson}. 
For simplicity,  we neglect the $\omega$-dependence of Im $\epsm$ and consider 
Im $\epsm$ = 8000, 18000 and 38000 for the three cases presented in Figure 2.
For each case, we present results of transmission for different number 
(1 to 10) of unit cells. Notice that the higher imaginary part of 
the metal the higher is the transmission. Also the losses due to the absorption 
are smaller, as can be seen from the decrease of the transmission peak as 
the length of the system increases. This result is
consistent with the formula presented by Pendry et al. \cite{ieee}
 for the effective permeability of the system
\be\label{muef}
\mu_{\rm eff}=1-\frac{F\nu^2}{\nu^2-\nu_0^2+i\gamma\nu}
\ee
with the damping factor
\be\label{gamma}
2\pi\gamma=\frac{2L_x\rho}{r\mu_0}
\ee
and the resonance frequency
\be\label{omega}
(2\pi\nu_0)^2=\frac{3L_x\cl^2}{\pi\ln\left(\frac{2c}{d}\right)r^3}.
\ee
where $\rho$ is the resistance of the metal, $L_x$ is the size of the system 
along the $x$ axis, $\cl$ is a the velocity of light in vacuum  and parameters
$r$, $c$ and $d$ characterize the structure of SRR. They are defined in Figure 1b.
Notice that the damping term $\gamma\to 0$ as $\sigma\to\infty$.
Since the Im $\epsm$ is proportional to $\sigma$,  $\gamma$ is inversely proportional 
to Im $\epsm$. Our numerical results suggest that it is reasonable to expect that the LHM effect will be more pronounced in systems with higher conductivity.

In Figure 3, we present the frequency dependence of the transmission for SRRs 
with the same parameters as those in Figure 2. Notice that the transmission 
is more pronounced as the length of the system is increased. 
Note also that the resonance gap becomes narrower when Im $\epsm$ increases. 
This  is in agreement with  Eq. (\ref{muef}). The frequency interval, in which
the effective permeability is negative, becomes narrower when the damping
factor $\gamma$ decreases.

In Figure 4 we show  the transmission through the  LHM, in which  the 
SRR are turned around their axis by 90 degrees. 
If we keep the same size of the unit cell as that of Figure 2, 
we do not obtain any LHM peak in the transmission, 
although there is a very well defined gap for the SRR alone. 
It seems that for  this orientation of the SRR 
there is no overlap of the field of the wire with that of the SRR. 
The results shown in Figures 4 and 5 are therefore obtained with a reduced unit cell
of $3.66\times 3.66\times 3.66$ mm (the size of SRR is still $3\times 3$ mm). 
The LHM  transmission  peak is located 
close to the lower edge of the SRR gap (shown in Figure 5). 
This is in contrast to the results presented in Figures 2 and 3, 
where the LHM transmission peak is always located close to the upper edge of 
the SRR gap. 
Finally, the gap shown for the ``turned'' SRR shown in the Figure 5,
 is deeper and broader than the gap for the ``up'' SRR.

In fact, for the SRR ``up'' structure, we found that the transmission 
in the gap is always of order of $10^{-7}-10^{-8}$.
We can explain this effect by non-zero transmission
from the $p$ to $s$ polarized wave (and back). If the transmission
$t(p\to s)$ and $t(s\to p)>0$, then there is always the 
non-zero probability $\propto t(p\to s)t(s\to p)$ 
for the $p$-polarized wave to switch into the $s$ state,
at the beginning of the sample, move throughout the sample as the $s$ wave
(for which neither wires nor SRR are interesting), and in the last unit cell
to switch back into the $p$-polarized state.  This process contributes to the 
transmission probability $T(p\to p)$ of the whole sample and  determines the bottom
level of the transmission gap for SRR. We indeed found that
$t(s\to p)\sim 10^{-4}$ for the ``up'' SRR. 
In the ``turned'' SRR case,  both $t(p\to s)$ and $t(s\to p)$ should be zero 
due to the symmetry of the unit cell.\cite{com} 
Our data give $t(p\to s)\sim t(s\to p)\sim 10^{-6}$ for the ``turned'' SRR
which determines the decrease of the transmission in the gap below 
$10^{-11}$.

\subsection{Dependence  on the size of the unit cell and the width of the metallic wire.}\label{3b}

As we discussed in Section \ref{3a}, the  transmission peak for the LHM with 
cuts in  the SRR in the horizontal direction appears  only 
when the size of the unit cell is really small. 
The effect of the size of the unit cell was demonstrated already in Figures
2 and 3, where we compared the transmission for the ``up'' SRR and LHM of different
size of the unit cell. Both the transmission gap for an array of SRRs and the transmission peak for LHM are broader for smaller unit cell. 

In Figure  \ref{ex}, we  show the transmission for the LHM structure
with q  unit cell 
of size $5\times 3.66\times 5$ mm for the  ``turned'' SRR. 
Evidently, there is no transmission peak for all the system lengths studied. 
The size of the unit cell must be reduced  considerably to
obtain a transmission peak. 

Figure \ref{fss} presents the transmission peak for various  sizes of the unit cell.
Resonance frequency decreases as the distance between the SRRs in the $x$ direction
decreases.  This agree qualitatively (although not quantitatively) with theoretical
formula given by Eqn. (\ref{omega}).
We see also that an increase of the distance between SRR in the $z$ direction 
while keeping the $L_x$ constant
causes sharp decrease and narrowness of the transmission peak.

\subsection{Resonance frequency of SRR}\label{3c}

In this section we study how the structure of the SRR influences
the position of the resonance gap.  
In order to simulate various forms of the SRR, we need to have as many as possible
mesh points in the $yz$ plane. Keeping in mind the increase of the computer time when
the number of mesh points increases, we used
a unit cell with $L_x<L_y,L_z$. 
The actual size of the unit cell in this section is
$L_x=2.63$ mm and  $L_y=L_z=6.05$ mm.  
and we use  uniform discretization  with 
$N_x\times N_y\times N_z=10\times 23\times 23$ mesh points. 
This discretization defines a minimum unit length $\delta=0.263$ mm. 
SRR with  size of  $\approx 5\times 5$ mm, is divided into  $19\times 19$ mesh points.

The electrical permittivity of the metallic components is chosen to be
$\epsm = -1000 + 10.000~i$. 
We  expect that 
larger value of Im $\epsm$ will increase a little  the position of the resonance gap
\cite{ms}. However, the dependence on the different structural parameters will remain the same.
Higher  values of $|\epsm|$, however,
will require more CPU time because of shorter interval between 
the normalization of the transmitted waves (see Appendix A for details).

We have considered 23 different SRR structures and studied how the
resonance frequency $\nu_0$ depends on the structure parameters.
The lowest $\nu_0=3.75$ GHz was found for SRR with c~:~d~:~r~:~g~=~1~:~1~:~13~:~1.
On the other hand, SRR with c~:~d~:~r~:~g~=~2~:~3~:~5~:~3 exhibits $\nu_0=6.86$ GHz.

We present our results on the dependence of $\nu_0$ on the azimuthal gap  $c$ (Figure
\ref{azimuthal}), radial gap $g$ (Figure \ref{radial}) and ring thickness $d$ (Figure
\ref{thickness}). In all these cases $\nu_0$ increases as the different
parameters increase. The dependence shown in Figures \ref{azimuthal}-\ref{thickness}
agrees qualitatively with those done by a different numerical method \cite{others}.
When compare our results with 
the analytical arguments presented by Pendry et al. 
\cite{ieee}, we have to keep in mind that that various assumptions about the 
structural parameters have been done in derivation of Eqn. (\ref{omega}), which are not 
fulfilled for our structure.  
Note also that the azimuthal gap does not enter the formula 
for the resonance frequency given by Eq. (\ref{omega}).
Moreover, as the  size of the SRR is constant, the structural parameters are
not independent each form other. Thus, due to the  Eq.  (\ref{ww}), 
increase of the azimuthal gap causes decrease of the inner diameter
and vice versa. 
When taking these restrictions into account, the agreement with analytical
results  is  satisfactory.

\subsection{Material parameters}\label{3d}

In Figures \ref{board} and \ref{air} we show how the resonance frequency
depends on the permittivity of the dielectric board and on the permittivity 
of the embedding media. As expected,  the resonance frequency decreases considerably
with the increase of the value of both  permittivities.

\section{Conclusion}\label{4}

In summary, we have used the transfer matrix method to calculate the 
transmission properties of the left-handed materials and arrays of split 
ring resonators. The role
of absorption of the metallic components of our SRR and LHM has been 
simulated. It is found that the LHM transmission peak depends on the 
imaginary part of the metallic
permittivity $\epsm$, the length of the system and the size of the unit cell. 
Higher conductivity  of the metal guarantees better transmission properties of LHM.

For an array of SRR, the resonance frequency $\nu_0$ was computed and is found 
to agree with experimental data.  The dependence of the resonance frequency 
$\nu_0$ on various structural parameters of the SRR were numerically obtained 
and compared  with analytical estimates and also with other numerical 
techniques.  

The main disadvantage of the presented transfer-matrix method is that
it can not treat structures with smaller length scales than our discretization mesh.
For example,  the thickness of the SRR
is an order of magnitude smaller in experiments than in our simulation.
Also structural parameters of SRR can be changed only discontinuously as 
multiplies of the unit mesh length.  This could be partially overcomed by
generalizing  the present code to a non-uniform mesh discretization.
Nevertheless, already  uniform discretization enables us to obtain credible
data. Comparison of our results with
those obtained by the commercial software MAFIA \cite{dv} confirmed that both
methods find the same position of the resonant gap provided that they use
the same mesh discretization. 

Our numerical data  agree qualitatively with the experimental results.
\cite{padilla} As we can not tune the exact parameters of SRR
(as well as its circular shape), and when taken into account the strong
dependence of the resonance frequency on the permittivity of the board,
we do not expect to obtain very accurate quantitative agreement with
experimental data.

Our studies demonstrate that the transfer matrix method can be 
reliable used to calculate the transmission and reflection properties 
of left-handed materials and split-ring resonators.
Thus, numerical simulations could answer some practical questions about different
proposed structures, which might be too complicated to be treated by
analytical studies. The transfer matrix method 
 can be used in the future for detailed studies  of two-dimensional
and even three-dimensional structures. These structures
should contain more  SRRs and wires per unit cell, which makes their analytical
analysis extremely difficult. On the other hand,
it is extremely important to find the  best design and test the transmission
properties of proposed meta-material even  before  their fabrication and experimental
measurements start.

%%%%%%%%%%%%%%%%%%%%%%%%%%%%%%%%%%%%%%%%%%%%%%%%%%%%%%%%%%%%%%%%%%%%

%\section{Appendix}
\appendix\section{}

Transfer matrix calculations are based on the scattering formalism.
The sample is considered as the scatterer of an incoming wave. 
The  wave normalized to 
the unit current is coming from the $-\infty$, and is scattered by the sample.
Scatterer is characterized by four parameters: transmission of the wave
from the left to the right ($t_+$), from the right to the left ($t_-$),
and by reflection coefficient from the right to the right ($r_+$) and from the left
to the left ($r_-$). Corresponding  scattering matrix $S$ reads:
\be\label{one}
S=\left(\begin{array}{ll}
		t_+ &  r_+\\
		r_- &  t_-
	\end{array}\right)
\ee
which determines the amplitudes of the outgoing waves $B,C$ in terms of 
the amplitudes of the incoming waves $A,D$:
\be\label{two}
\left( C\atop B\right)
=
S
\left( A\atop D\right)
\ee
Relation (\ref{two}) can be re-written into the form
\be\label{twox}
\left( D\atop C\right)
=
{\cal T}
\left( B\atop A\right)
\ee
${\cal T}$ is the transfer matrix, which determines the fields on one side
of the sample with the fields on the another side.
Its explicit form reads
\be\label{TM}
{\cal T}=\left(\begin{array}{ll}
t_-^{-1} &
 -t_-^{-1}r_-   \\
 r_+t_-^{-1} &
t_+-r_+t_-^{-1}r_-\\
\end{array}\right).
\ee

Transfer matrix (TM) fulfills the composition law. If the sample consists from two
subsystems, then the transfer matrix ${\cal T}_{12}$ of the whole sample can 
be calculated  form transfer matrices of its subsystems as
\be\label{cl}
{\cal T}_{12}={\cal T}_2{\cal T}_1.
\ee
Resulting TM ${\cal T}_{12}$ has again the form (\ref{TM}). This composition law
enables us to calculate transmission of complicated structure from the transfer
matrices of its parts (thin slices).

In numerical calculations, the total volume of
the system is divided into small cells and fields in each cell are
coupled to those in the neighboring cell. 
We discretize the Maxwell equations following the method described in Ref. 13.
In each point of the lattice we have to
calculate four components of the EM field: $E_x$, $E_y$, $H_x$, $H_y$.

We assume that our system is connected to two semi-infinite leads
(with $\eps=1$ and $\mu=1$). EM wave is coming from the  right
and is scattered by the sample. Resulting waves either continues
to the left on the left side lead, or are traveling back to the right
on the right side lead. 
Periodic boundary conditions in the directions perpendicular to the 
direction of the wave propagation are used.

We decompose the system into $n$ thin slices and define a TM for each of them.
Explicit form of the TM for a thin slice is in Refs. 12 and 13. 
The EM field in the $(k+1)$th slice can be obtain from the $k$th slice
as
\be
\Phi_{k+1}={\cal T}_k\Phi_k.
\ee
with ${\cal T}_k$ being the transfer matrix corresponding to  the $k$th   slice.
The transfer matrix  ${\cal T}$ of the whole sample reads
\be\label{iter}
{\cal T}={\cal T}_{n}{\cal T}_{n-1}\dots{\cal T}_2{\cal T}_1 
\ee

If there is $N$ mesh points in the slice, then the length of the vector $\Phi$  
is $4N$ (it contains 4 components of EM field in each point).
%\cite{122,pendry}.

Note that we are able to  find the explicit form of the  TM only in the real space
representation. 
To obtain the transmission, we have to transform the TM 
into the ``wave'' representation, which is defined  by the
eigenvectors of the TM in the leads. 
Therefore, in the first step we have to  diagonalize  the TM in the leads.

Each eigenvalue of the TM is two time degenerate because there are
 two polarizations $p$ and $s$ of the EM wave. Moreover,
if $\lambda$ is an eigenvalue, then $\lambda^{-1}$ is also an eigenvalue
corresponding to the wave traveling in the opposite direction.
In general, the TM has some eigenvalues with modulus equal to 1:
$\lambda=\exp ik$. The corresponding eigenvectors represent 
propagating waves. Others eigenvalues  are of the form
$\lambda=\exp \pm\kappa$.  They correspond to the evanescent modes. 
For the frequency range which is interesting for the LHM studies,
the TM has only one propagating mode.

As the TM is not Hermitian matrix, we have to calculate left and right
eigenvectors separately. From the eigenvectors we construct three
matrices: The $2N\times 4N$ matrix $R_1$ contains in its columns 
 $2N$ right eigenvectors
which correspond to the wave traveling to the left.
Matrices $L_1$ and $L_2$ are $4N\times 2N$ matrices which contains in their
rows the left eigenvectors corresponding to waves traveling 
to the left and to the right, respectively.

The general expression of  the TM given by Eq. (\ref{TM})
enables us to find the transmission matrix explicitly  \cite{12}
\be\label{ltr}
t_-^{-1}=L_1{\cal T}R_1
\ee
and the reflection matrix from the relation
\be\label{lrr}
r_+t_-^{-1}=L_2{\cal T}R_1.
\ee
At this point we have to distinguish between the propagating and the evanescent  modes. 
For a frequency range of interest, the TM in leads has only  one 
propagating mode for each direction. We need therefore only $2\times 2$
sub-matrices $t_-(ij)$ and $r_+(ij)$ 
with $i,j=1$ or 2 for the $p$ or $s$ polarized wave. 
The transmission and reflection are then
\be
T_{ij}=t_-(ij)t_-^*(ij)\quad\quad 
R_{ij}=r_+(ij)r_+^*(ij)
\ee
and absorption 
\be\begin{array}{ll}
A_{p} &=1-T_{pp}-T_{ps}-R_{pp}-R_{ps}\\
A_{s} &=1-T_{ss}-T_{sp}-R_{ss}-R_{sp}\\
\end{array}
\ee

It seems that relations (\ref{ltr}) and (\ref{lrr}) solve our problem completely.
However, the above algorithm must be modified. The reason is that
the  elements of the matrix $t_-^{-1}$ are given by their larger 
eigenvalues. 
We are, however,  interesting in the largest eigenvalues of the
matrix $t_-$.   
As the elements  of the transfer matrix increase
exponentially in the iteration procedure given by Eq. (\ref{iter}),
an information about the smallest eigenvalues of $t_-^{-1}$ 
will be quickly lost. We have therefore to introduce some re-normalization
procedure. We use the procedure described in
Ref. 16.

Relation (\ref{ltr}) can be written as
\be\label{new}
t_-^{-1}=L_1r^{(n)}
\ee
where we have defined $2N\times 4N$ matrices $r^{(k)}$, $k=0,1,\dots n$
as
\be\label{rr}
r^{(k)}={\cal T}_kr^{(k-1)},\quad\quad r^{(0)}=R_1
\ee
Each matrix $r$ can be written as 
\be
r=\left(r_1\atop r_2\right)
\ee
with $r_1$, $r_2$ being the $2N\times 2N$ matrices.
We transform $r$ as
\be\label{trik}
r=r' r_1\quad\quad r'=\left(1\atop r_2r_1^{-1}\right)
\ee
and define $r^{(k)}={\cal T}_k(r')^{(k-1)}$.
In contrast to $r_1$ and $r_2$, all eigenvalues of the matrix $r_2r_1^{-1}$
are of  order of unity.
Relation (\ref{new}) can be now re-written into the form
\be\label{kva}
t_-^{-1}=
L_1\left(1\atop r^{(n)}_2\left[r^{n}_1\right]^{-1}\right)
r^{(n)}_1r^{(n-1)}_1\dots r^{(1)}_1r^{(0)}_1
\ee
from which we get that 
\be\label{result}
\begin{array}{ll}
t_-= 
\left[r^{(0)}_1\right]^{-1}
\left[r^{(1)}_1\right]^{-1}\dots
\left[r^{(n)}_1\right]^{-1}
\left[ 
L_1\left(1\atop r_2^{(n)}\left[r^{(n)}_1\right]^{-1}\right)
\right]^{-1}.
\end{array}
\ee
From Eqn. (\ref{lrr}) we find
\be\label{resulr}
r_+=\left[L_2\left(1\atop r_2^{(n)}\left[r^{(n)}_1\right]^{-1}\right)\right]\times
\left[ 
L_1\left(1\atop r^{(n)}_2\left[r^{(n)}_1\right]^{-1}\right)
\right]^{-1}.
\ee

The   matrix inversion in the formulae (\ref{trik}-\ref{resulr})
can  obtained also by the soluton of a system linear equations.
Indeed,  matrix $BA^{-1}$ equals to matrix $X$, which solves the system of
linear equations  $B=XA$. CPU time could be reduces considerably in this way,
especially for large matrices.

All elements of the matrices on the rhs of Eqn. (\ref{result}) are 
of  order of unity. The price we have to pay  for  this stability is an 
increase of the CPU time. Fortunately, if the elements of the transfer
matrix are not too large (which is not the case in systems studied in 
this paper), then  it is enough to perform
described normalization procedure only after every 6-8 steps.

\bigskip

We thank D.R. Smith, M. Agio  and D. Vier for fruitful discussions.
Ames Laboratory is operated for the U.S.Department of Energy by Iowa
State University under Contract No. W-7405-Eng-82. This work was 
supported by
the Director of Energy Research, Office of Basic Science,
DARPA and NATO grant PST.CLG.978088. P.M. thanks
Ames Laboratory for its hospitality and support and Slovak Grant Agency
for financial support.

\begin{figure}
\epsfig{file=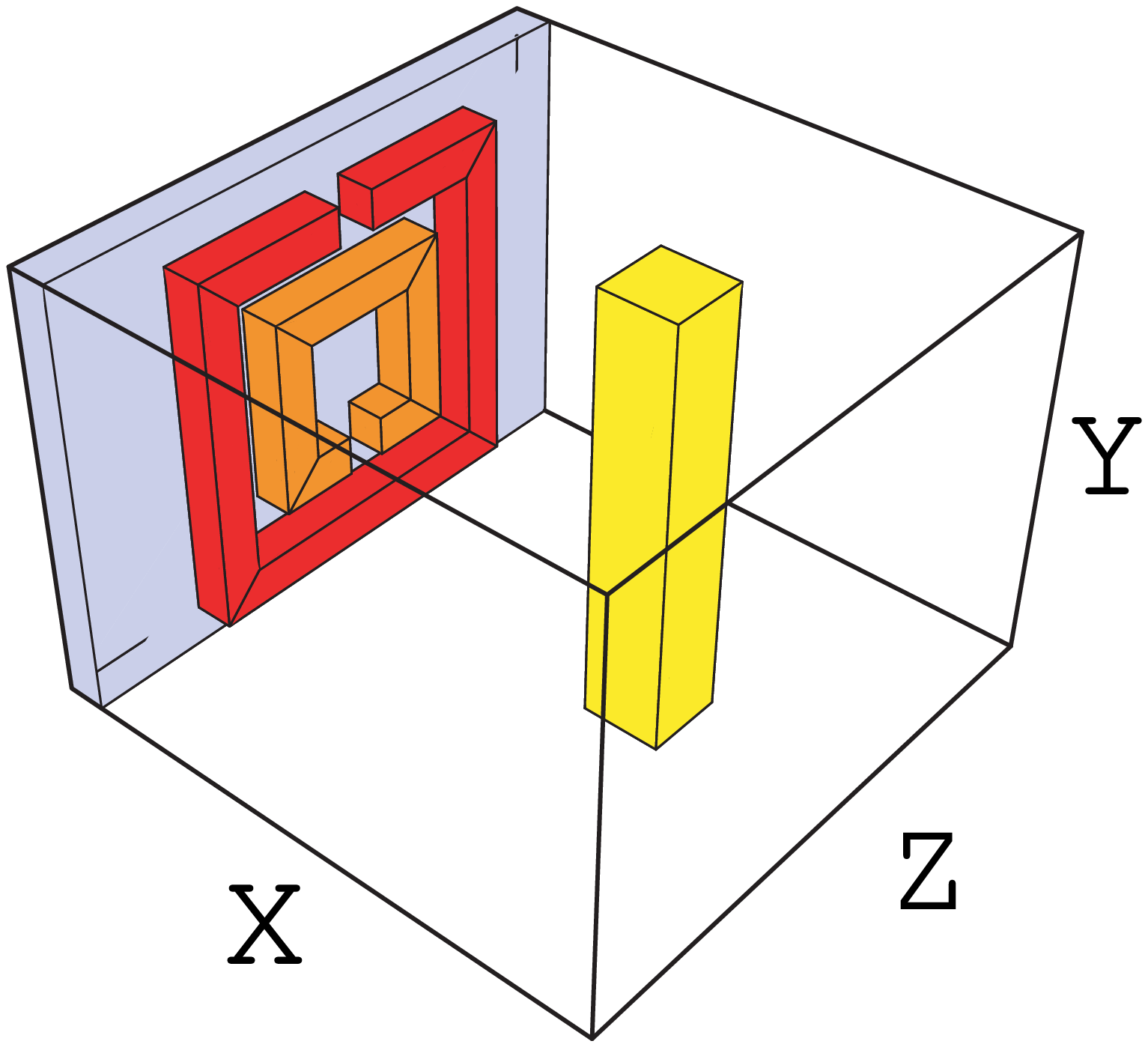,width=6cm}
\epsfig{file=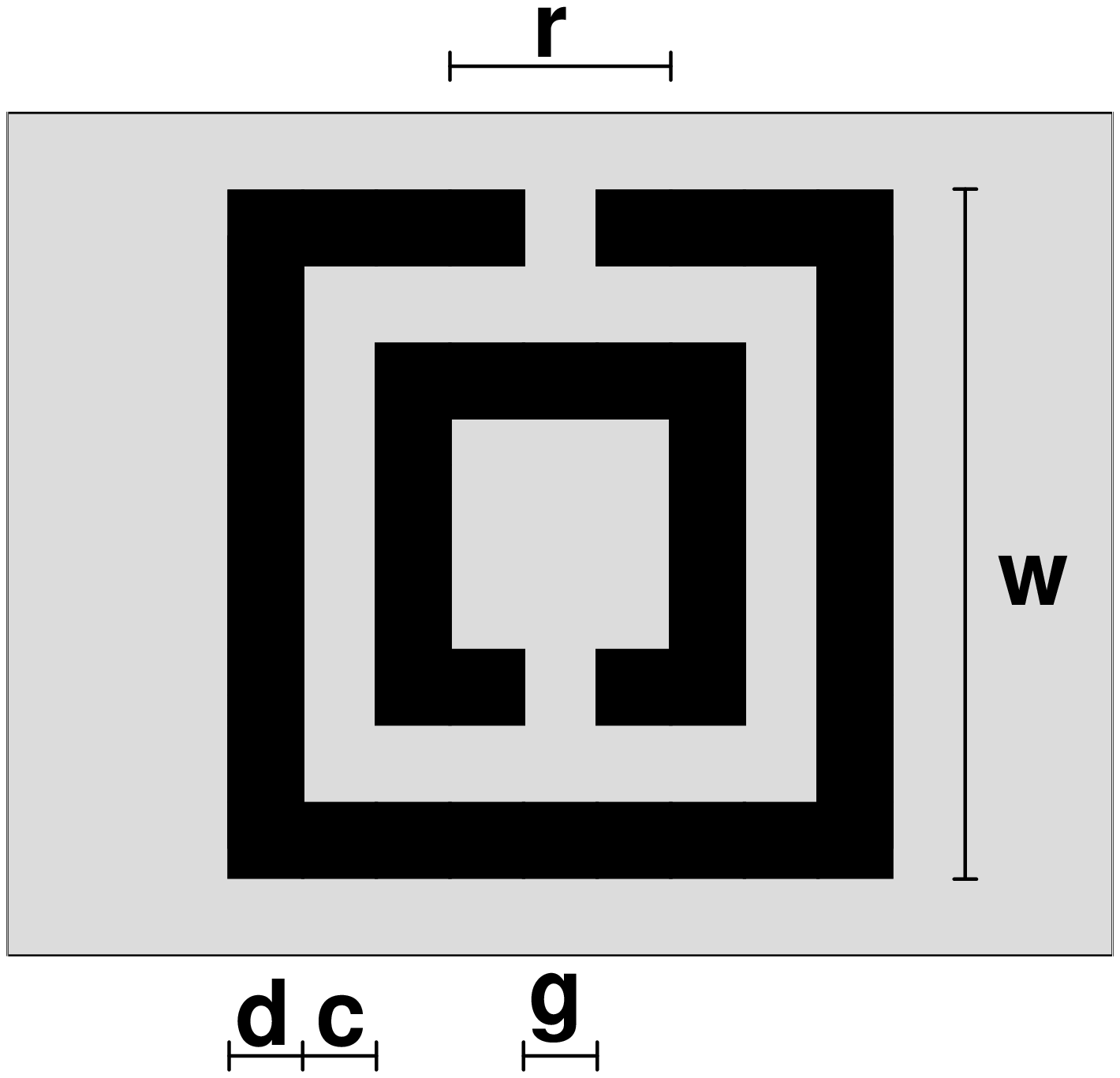,width=4cm}
\medskip
\caption{Top: The structure of the unit cell as was used in the present simulations.
Structure acts as the
left-handed meta-material if 
the electro-magnetic wave propagates along the $z$ direction and is polarized
with electric field {\bf E} parallel to the wire and magnetic field {\bf H}
parallel to the axis of SRR. Bottom:
The structure of the SRR and definition of the SRR parameters. 
}\label{structure}
\end{figure}

\begin{figure}
\epsfig{file=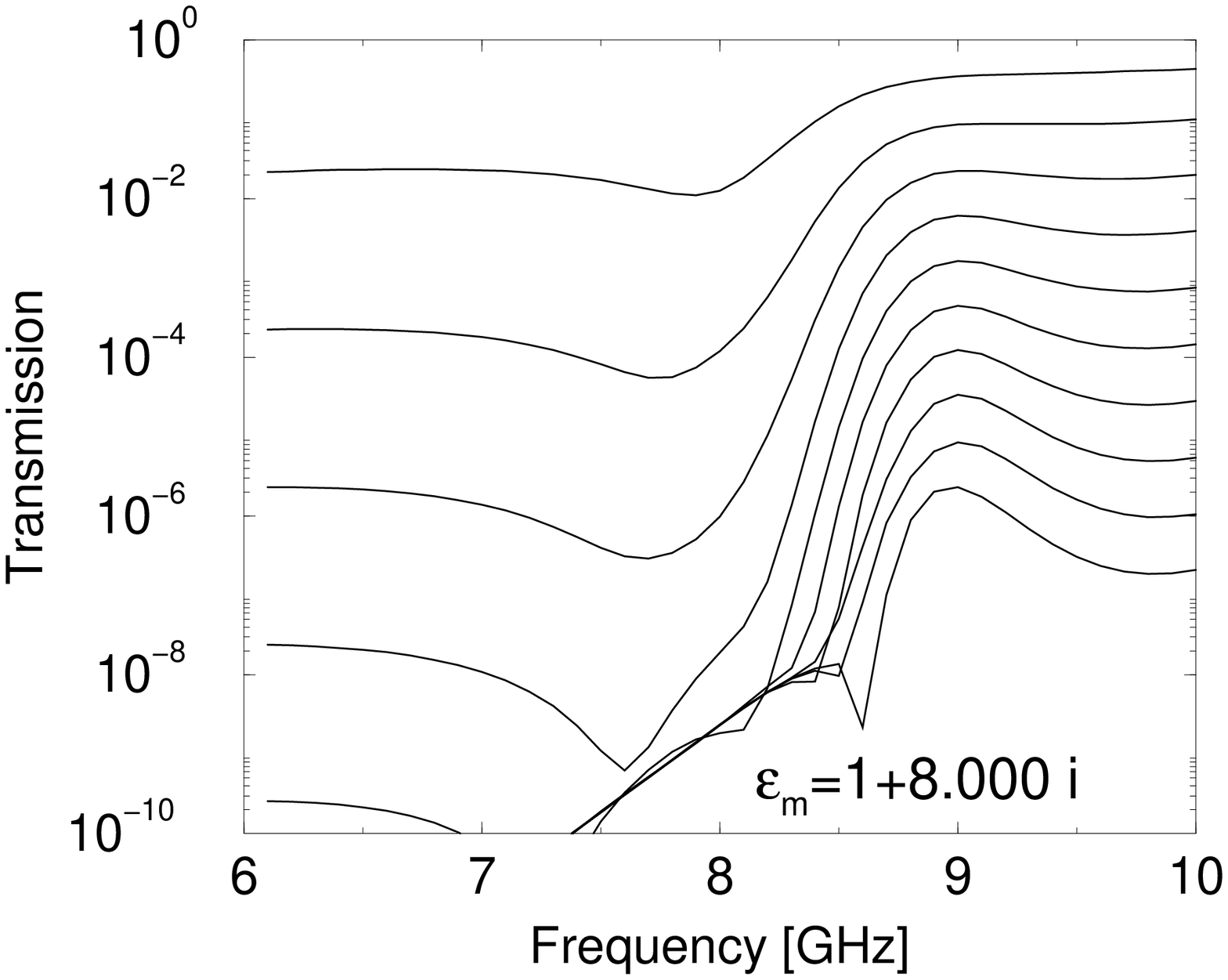,width=6cm}
\epsfig{file=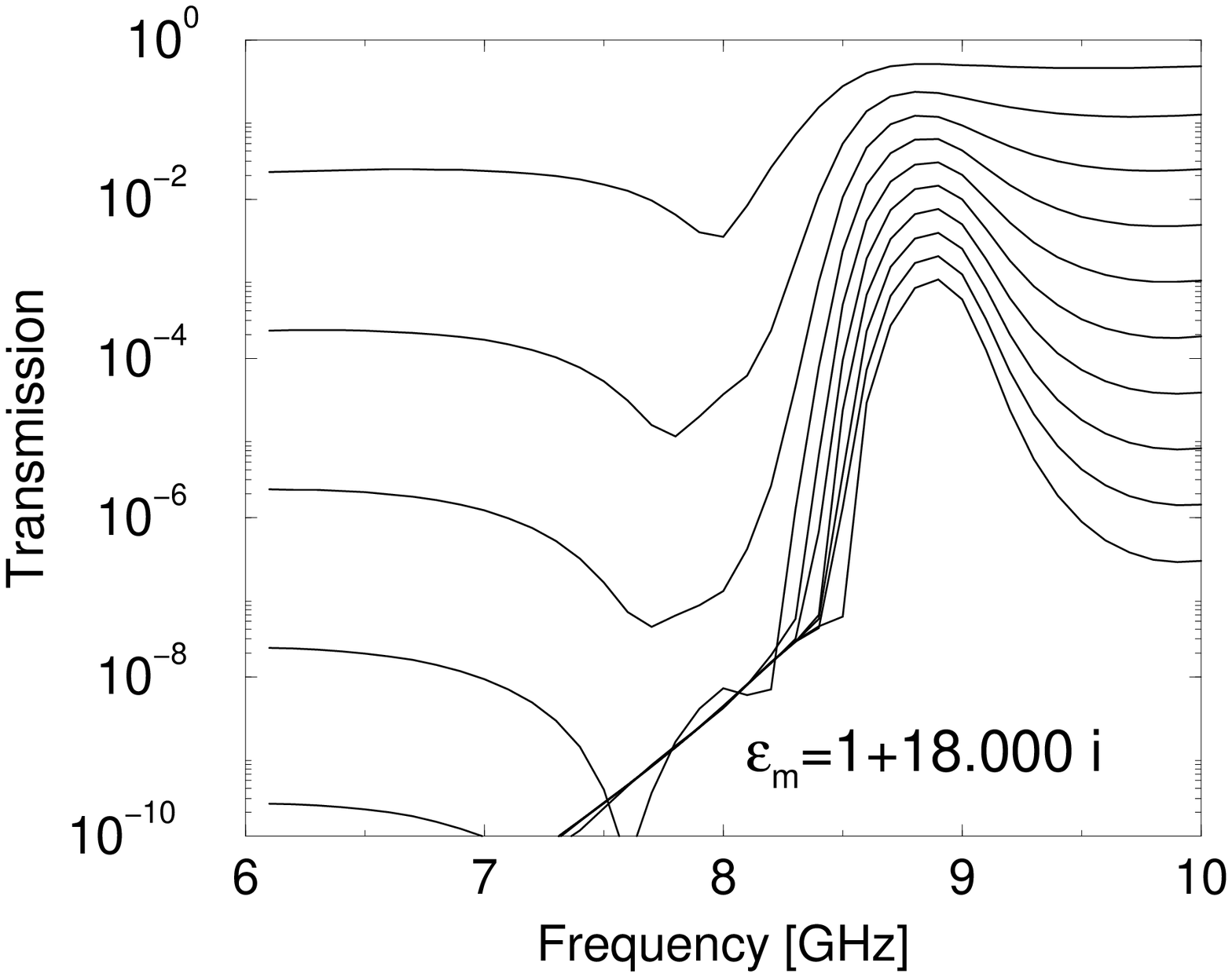,width=6cm}
\epsfig{file=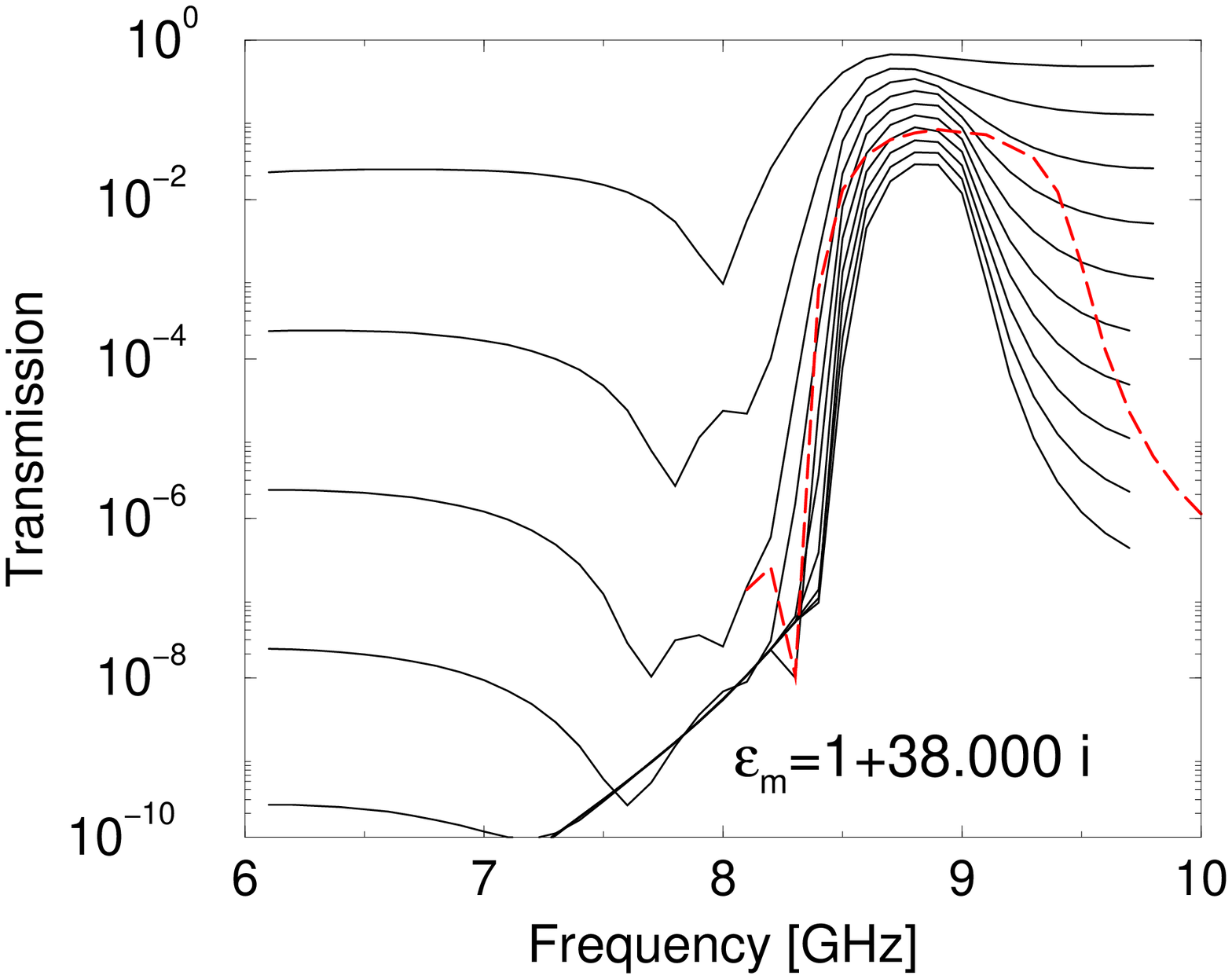,width=6cm}
\caption{LHM peak for various values of the metallic permittivity.
Lines corresponds to system length of 1,2,\dots 10 unit cells.
SRR is modeled as in Figure \ref{structure}. 
The size of the unit cell is $5\times 3.66\times 5$ mm,
the size of SRR is 3 mm, and the size of the wire is $1\times  1$ mm.
The dashed line is transmission for LHM system with unit cell $3.66\times 3.66\times 3.66$ and the system length of 10 unit cells.}
\label{metal_per}
\end{figure}

\begin{figure}
\epsfig{file=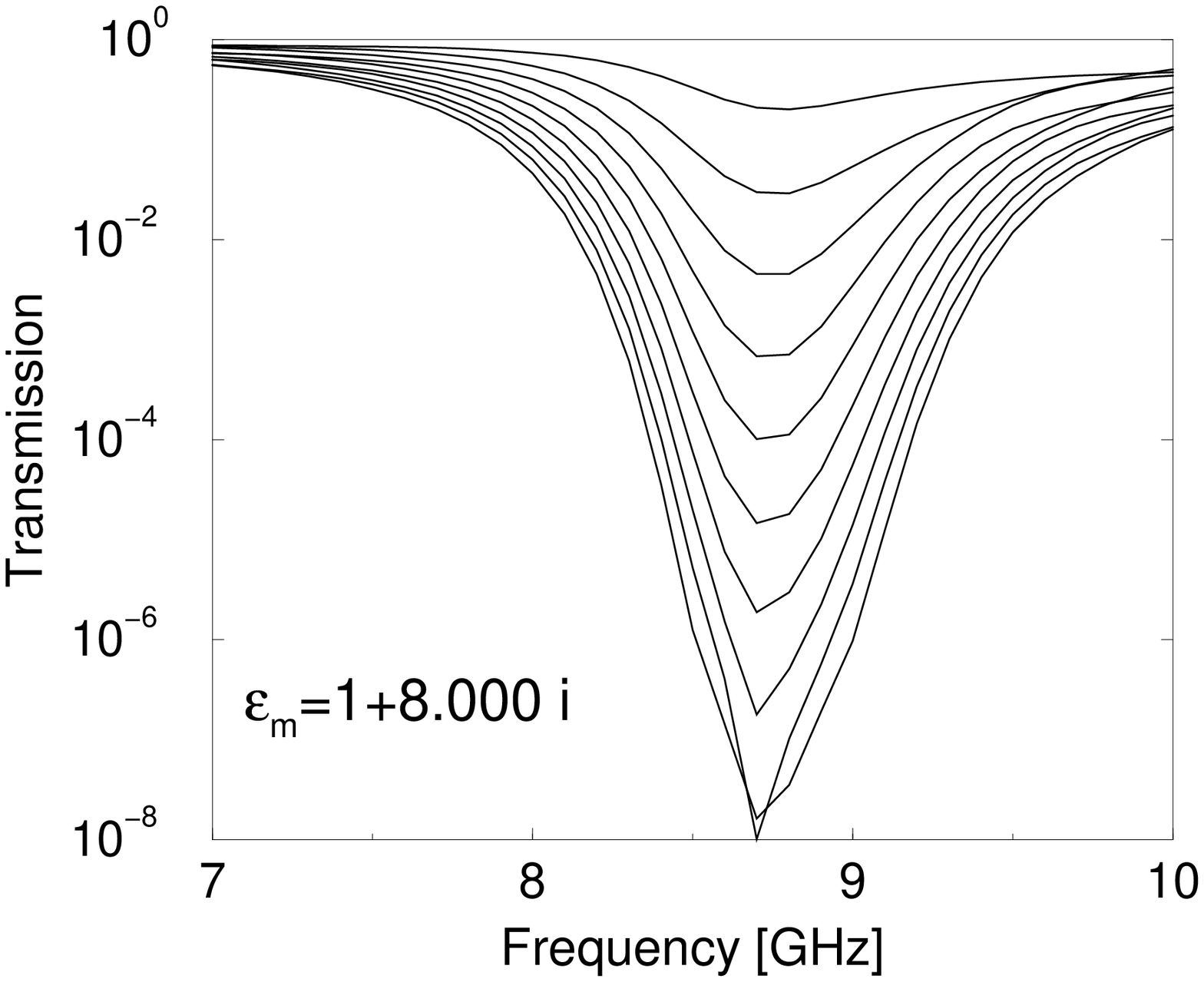,width=6cm}
\epsfig{file=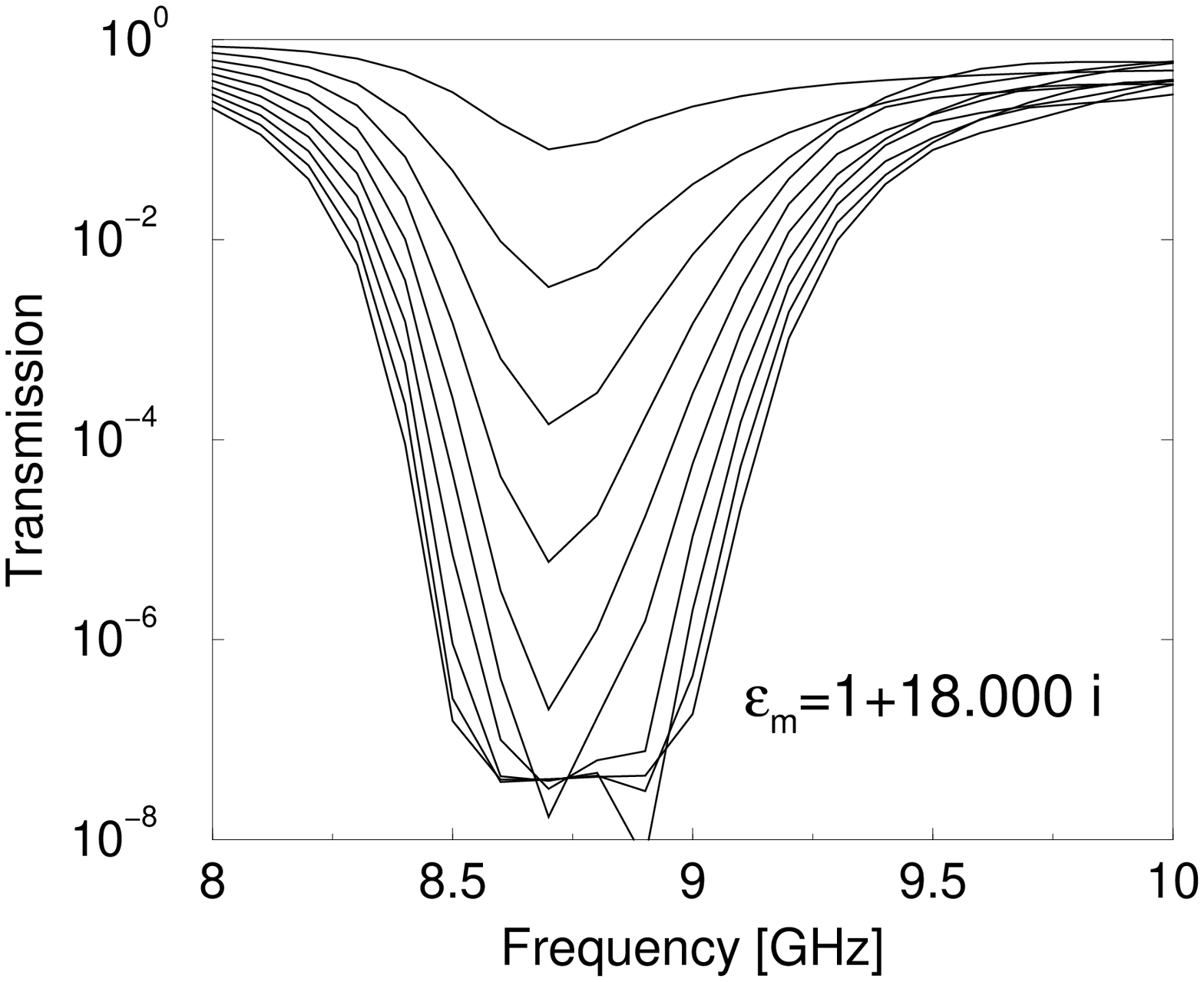,width=6cm}
\epsfig{file=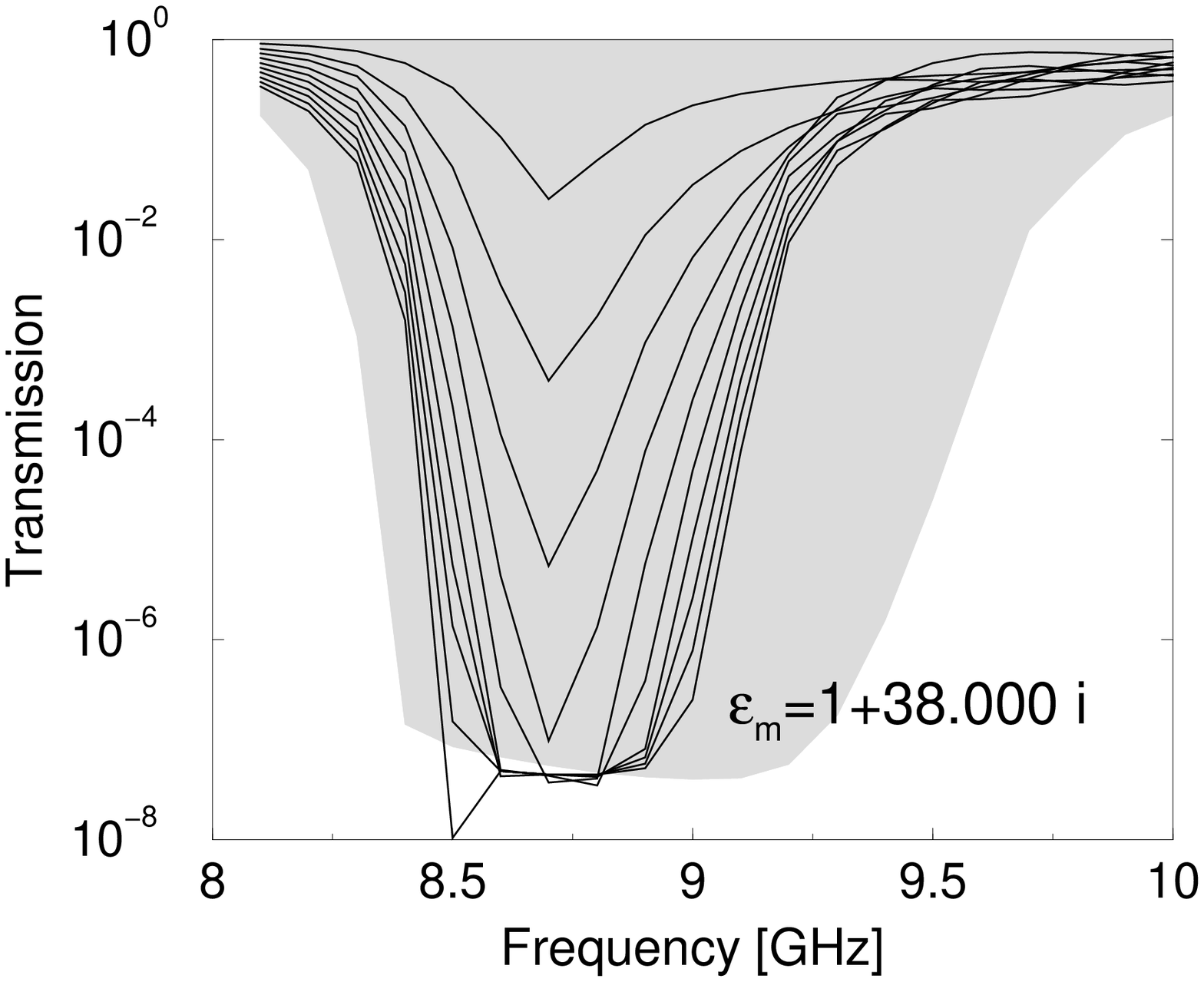,width=6cm}
\caption{Transmission for SRR for the same systems as in Figure \ref{metal_per}.
Data confirm that the resonance frequency does not depend on the  metallic 
permittivity. This agrees with \cite{ms}.
However, resonance gap becomes narrower as Im $\epsm$ of the 
metallic components increases.
Shaded area represents a gap for the array of the SRR with a unit cell
$3.66\times 3.66\times 3.66$ and  length system of 10 unit cells.
Note also that the  depth of the transmission  is constant ($\sim 10^{-7}$)
and see text for explanation.
}
\label{metal_per_srr}
\end{figure}

\begin{figure}
\epsfig{file=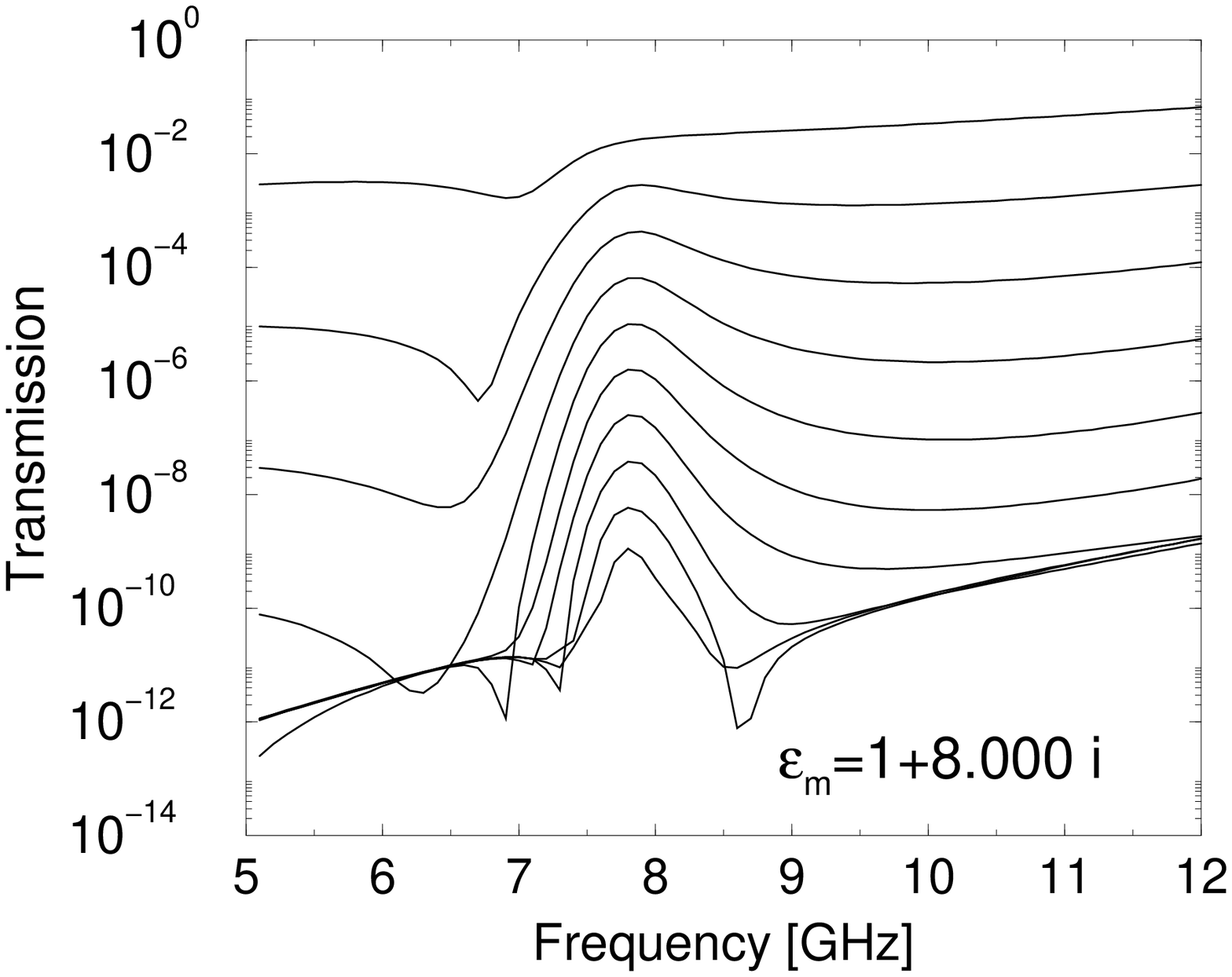,width=6cm}
\epsfig{file=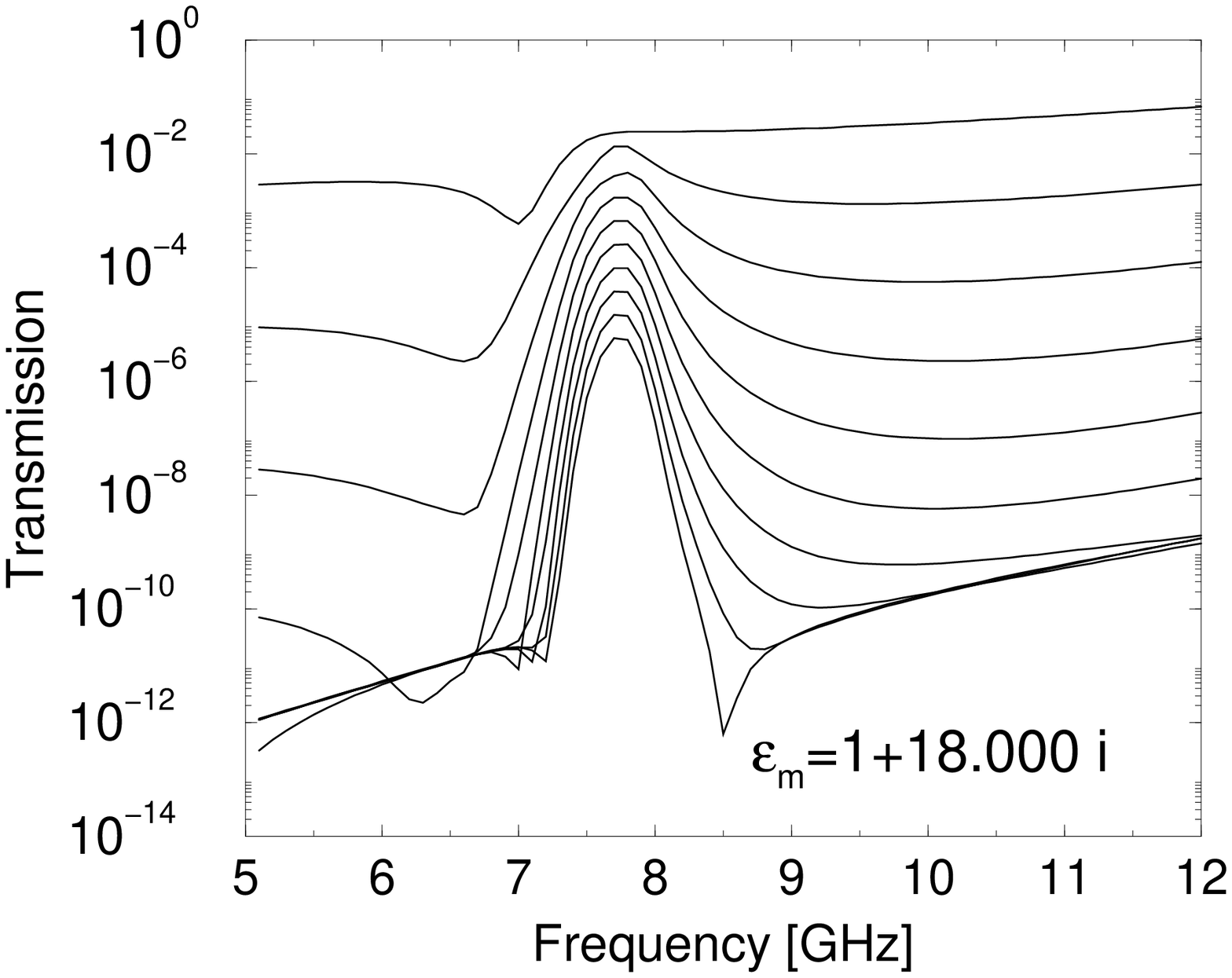,width=6cm}
\epsfig{file=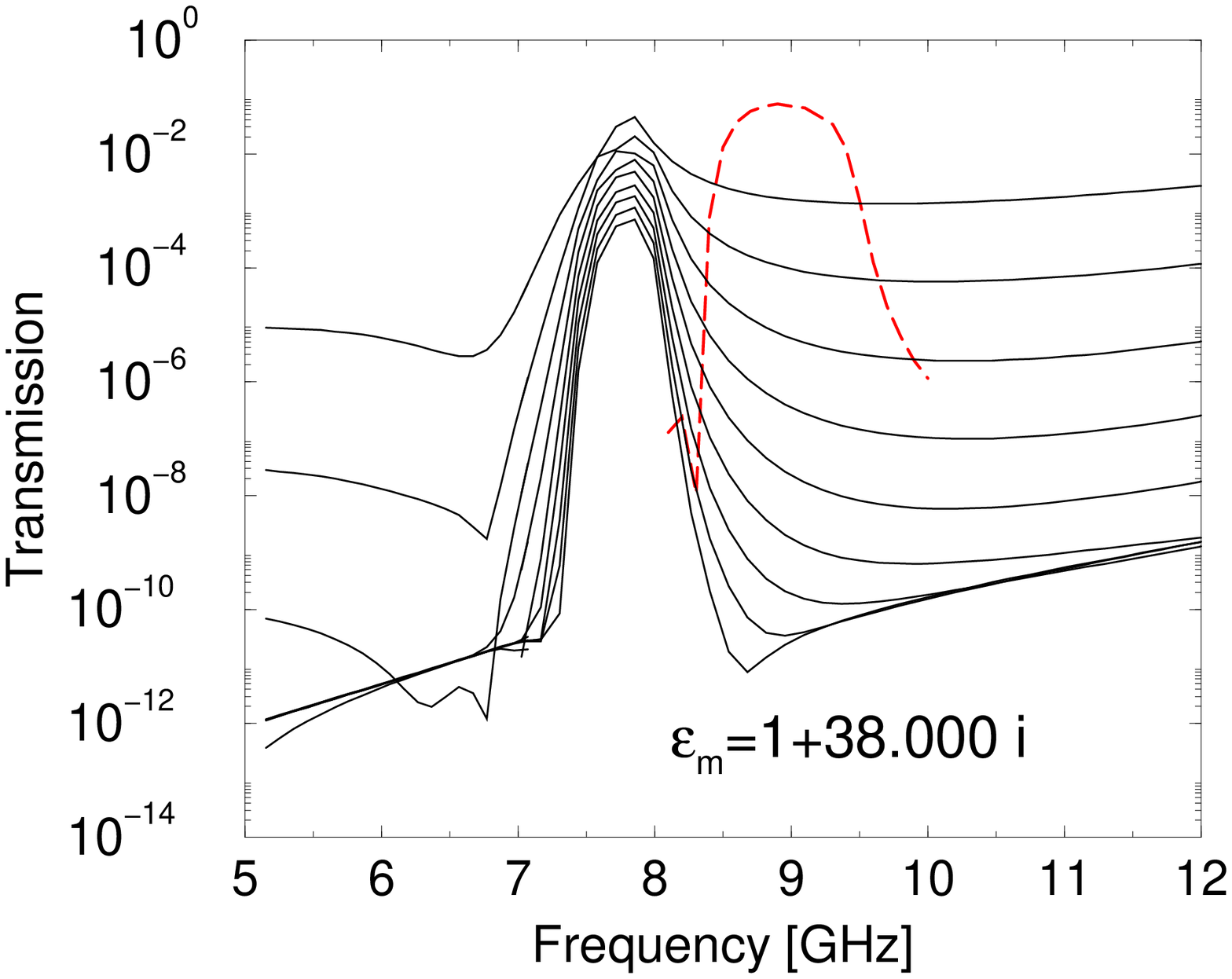,width=6cm}
\caption{Transmission for LHM with the SRR rotated by 90 degrees. 
The size of the unit cell is now $3.66\times 3.66\times 3.66$ mm.
We found no transmission peak for the unit cell size as in Figure \ref{metal_per}
(see Figure 11). 
For comparison with LHM with ``up'' oriented SRR, we show also the peak for this
structure of the same unit cell and metallic permittivity $\eps=1+38.000~i$
and length system of 10 unit cells 
(dashed line).
}

\end{figure}

\begin{figure}
\epsfig{file=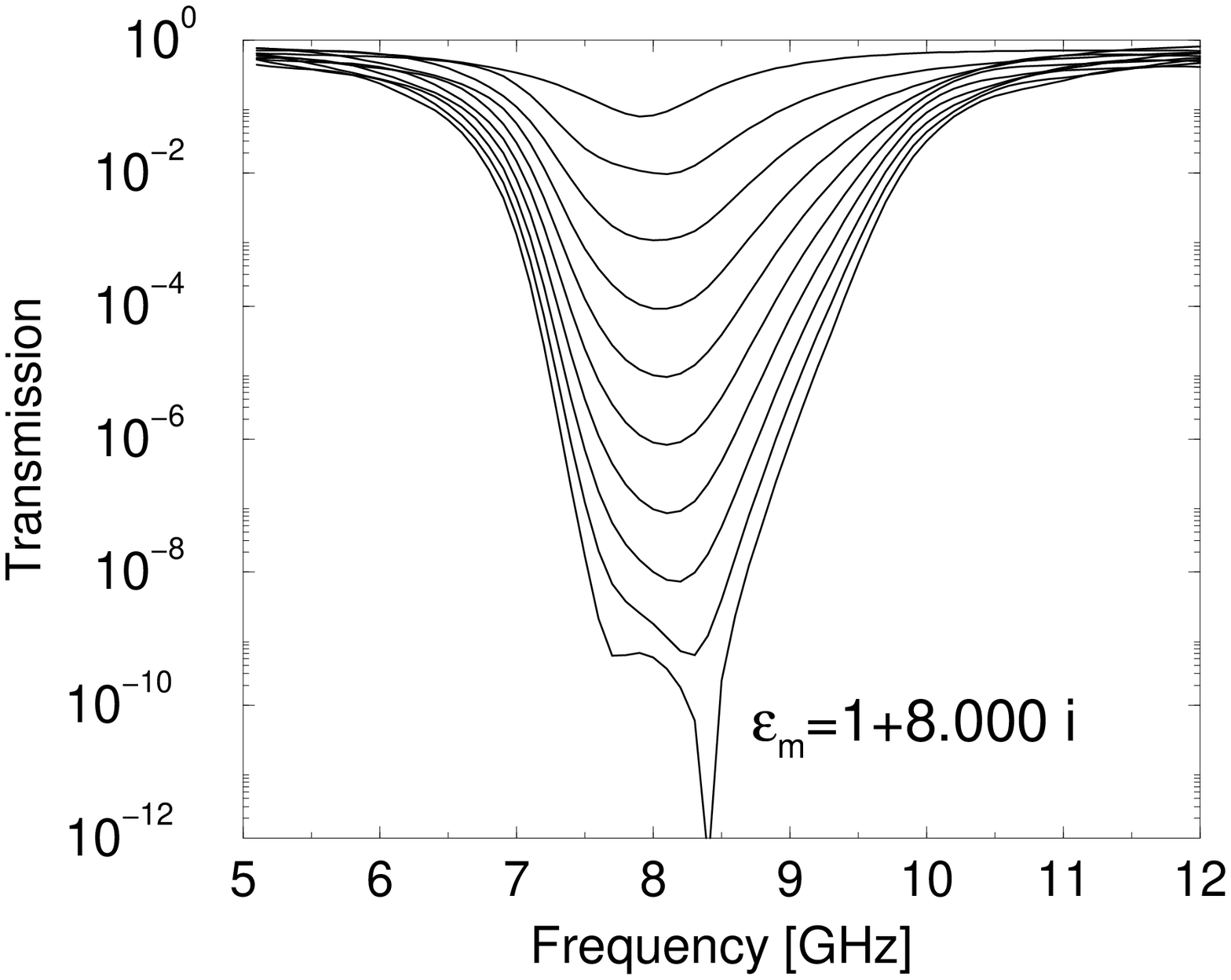,width=6cm}
\epsfig{file=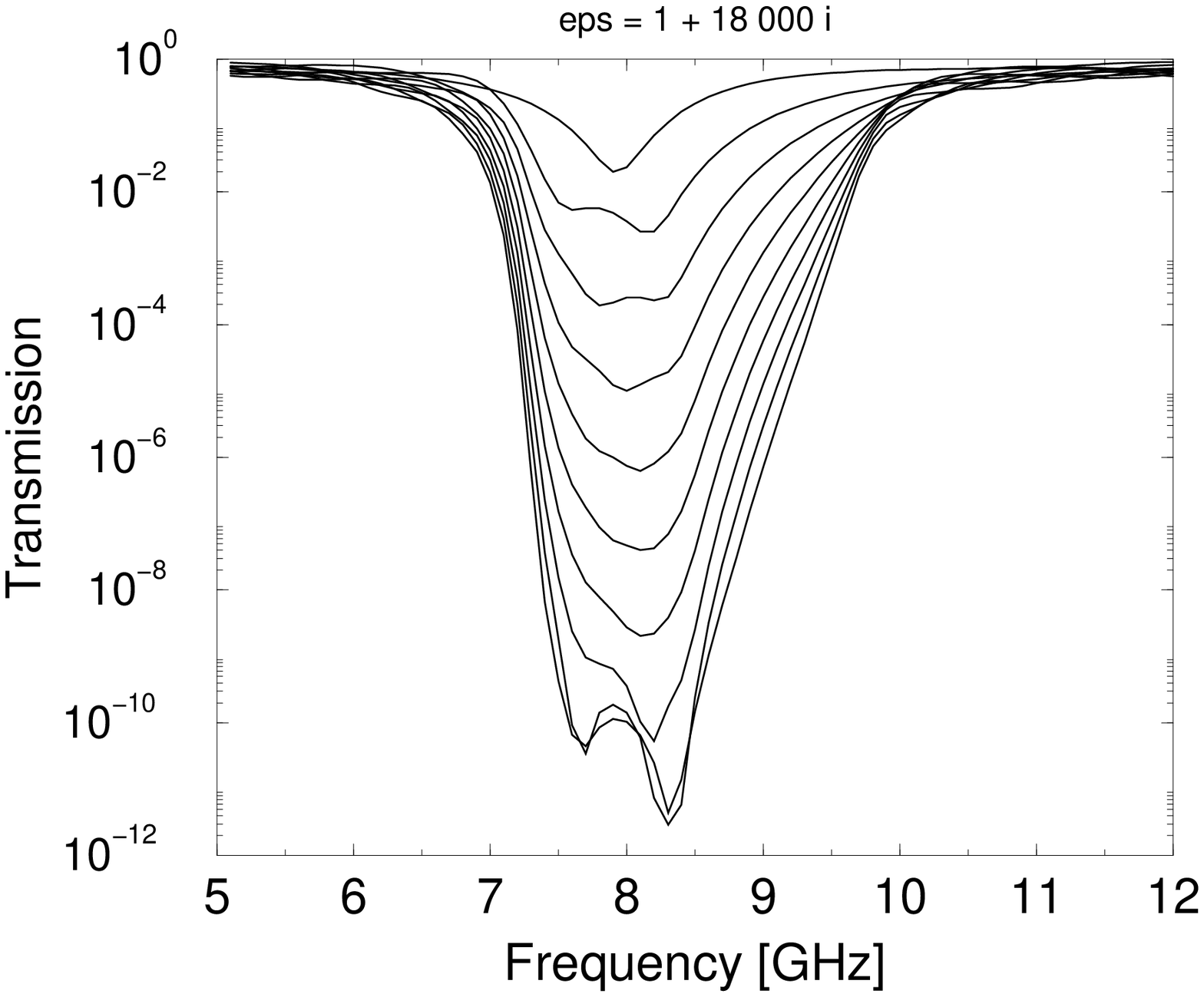,width=6cm}
\epsfig{file=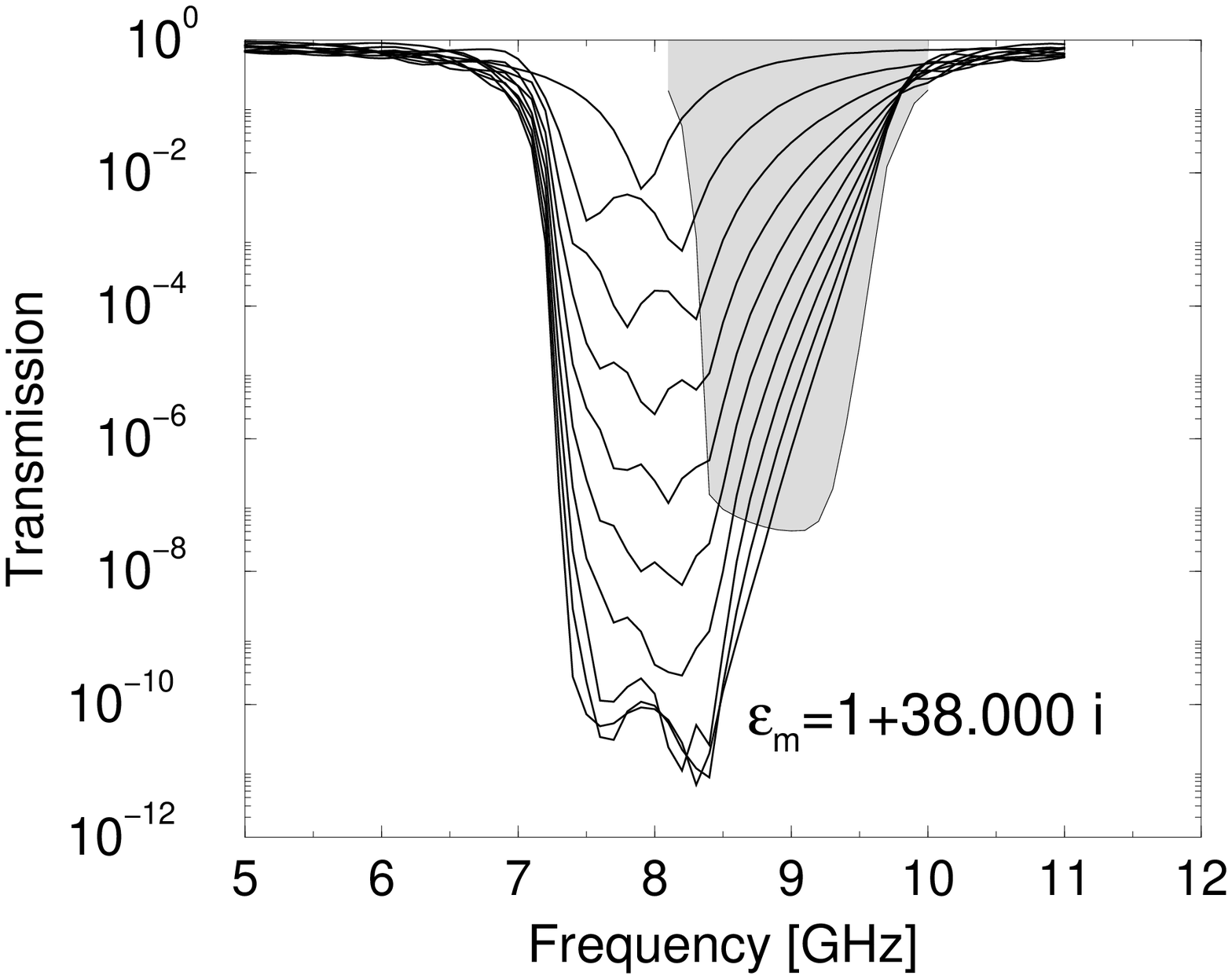,width=6cm}
\caption{Transmission for SRR rotated in 90 degrees. 
The size of the unit cell is now $3.66\times 3.66\times 3.66$ mm.
For comparison, we add also data for the array of  ``up'' SRRs and $\epsm=1+38.000~i$
and length system of 10 unit cells (shaded area). 
} 
\label{t}
\end{figure}

\begin{figure}
\epsfig{file=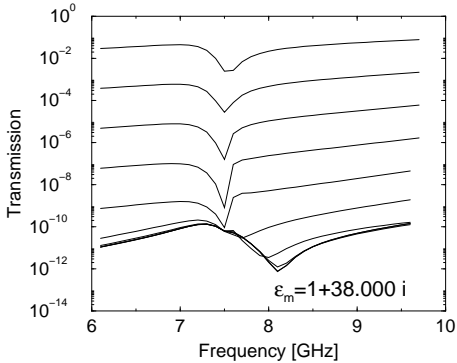,width=6cm}
\caption{The transmission for LHM for the same parameters as in Figure 4,
but with larger unit cell $5\times 3.66\times 5$ mm.  No transmission peak is
observed.}
\label{ex}
\end{figure}

\begin{figure}
\epsfig{file=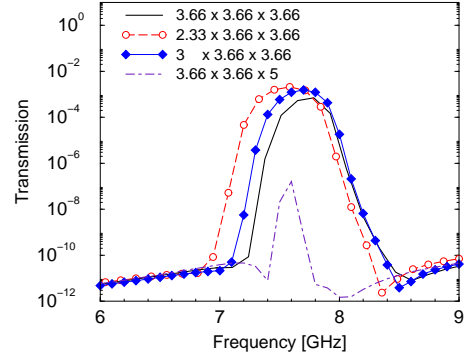,width=6cm}
\caption{Transmission peak for various sizes of the unit cell with ``turned'' SRR. 
The system length is 10 unit cells. Resonance frequency decreases slowly as the 
distance of SRR in the $x$-direction decreases. Increase of the distance
of SRR along the $z$ direction causes decrease of the transmission peak, which becomes also much narrower.}
\label{fss}
\end{figure}

\begin{figure}
\epsfig{file=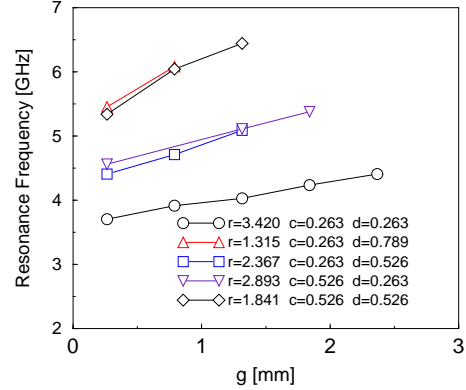,width=6cm}
\caption{%
Resonance frequency as a function of the azimuthal gap $g$
for various SRR structure. The
size of the SRR is 5 mm, the unit cell is $2.63\times 6.05\times 6.05$
mm, which corresponds to $10\times 23\times 23$ mesh points.
The ratio $d~:~c~:~r$ is 1~:~1~:~13, 3~:~1~:~5, 2~:~1~:~9, 1~:~2~:~11 and 2~:~2~:~7,
respectively. Note that azimuthal gap does not enter in Eqn. (\ref{omega}).
}\label{azimuthal}
\end{figure}

\begin{figure}
\epsfig{file=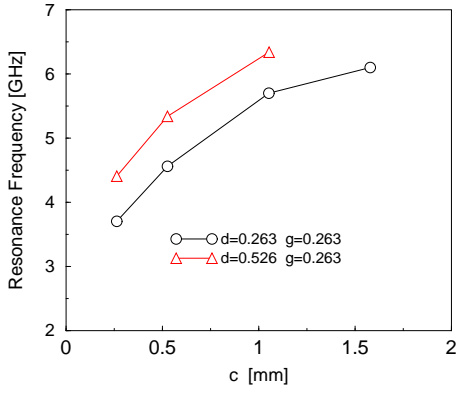,width=6cm}
\caption{%
Resonance frequency as a function of the radial gap $c$
(distance between rings) for various SRR structures.
The size of the sample and of the unit cell is an in Figure \ref{azimuthal}.
In contrast to Eqn. (\ref{omega}), which predicts decrease of $\nu_0$ when $c$
increases, we found increase of the resonance frequency. This could be
explained by the fact that
an increase of the radial gap causes decrease of the inner diameter
because $r+2c=3.947$ mm ($\circ$) and 2.895 mm ($\triangle$).
Presented data  can be therefore interpreted also as the inner diameter - 
dependence of the resonance frequency. The last ($\nu_0\sim r^{-3/2}$) is much stronger than the 
logarithmic dependence $\nu_0\sim \ln^{-1/2}c$.
Then, presented data confirm  that the  decrease of the inner diameter
causes an increase of the resonance frequency.  
}\label{radial}
\end{figure}

\begin{figure}
\epsfig{file=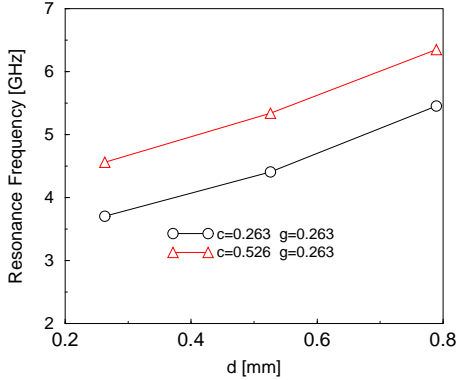,width=6cm}
\caption{%
Resonance frequency as a function of ring thickness for two sets of SRR.
The size of the sample and of the unit cell is an in Figure \ref{azimuthal}.
}\label{thickness}
\end{figure}

\begin{figure}
\epsfig{file=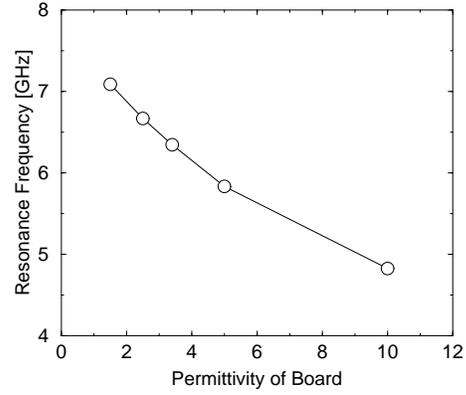,width=6cm}
\caption{%
Resonance frequency as a function of the permittivity of board.
The size of the sample and of the unit cell is an in Figure \ref{azimuthal}.
}\label{board}
\end{figure}

\begin{figure}
\epsfig{file=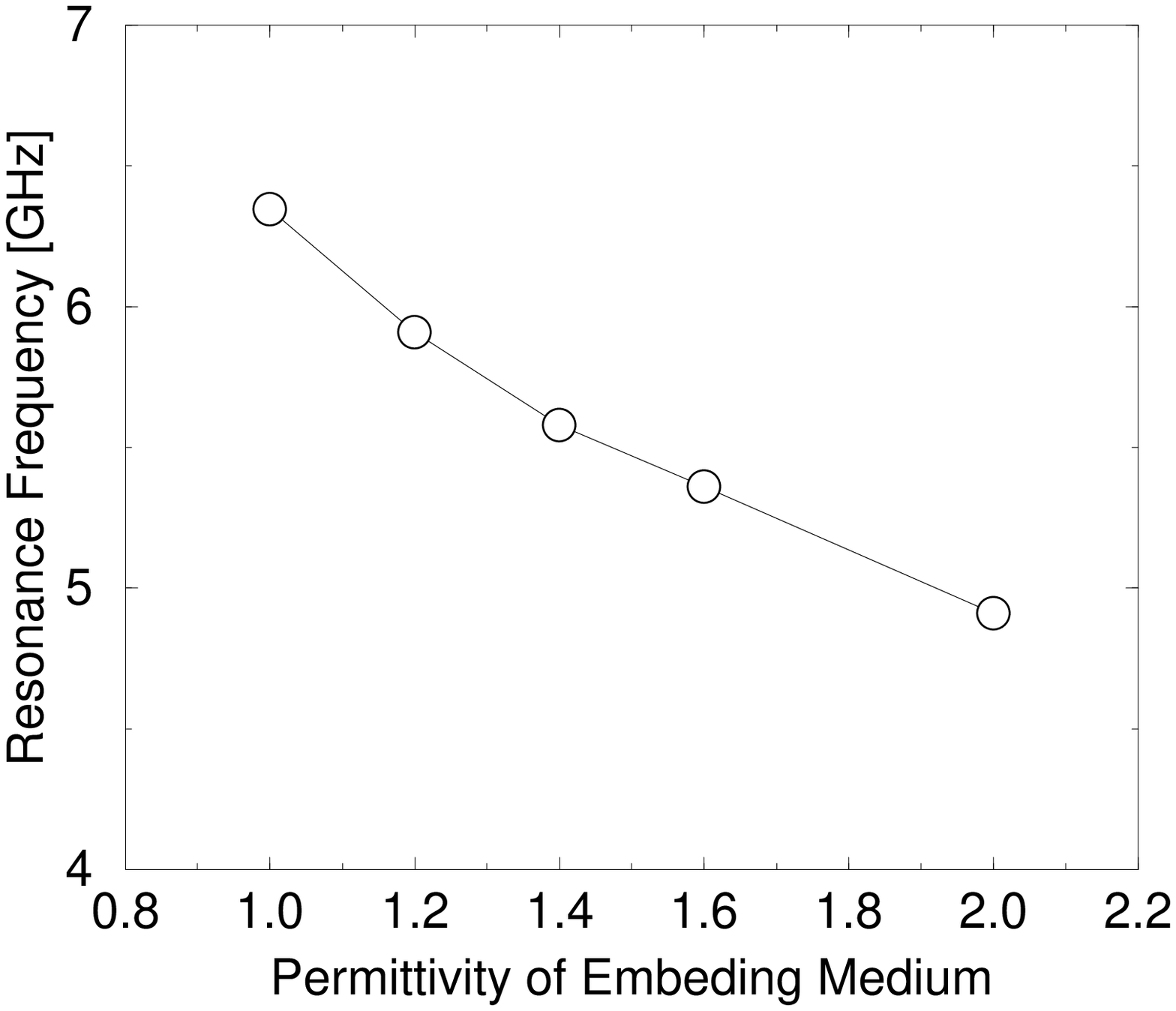,width=6cm}
\caption{%
Resonance frequency as a function of the permittivity of embedding media
in the unit cell (the permittivity of semi-infinite leads remains 1).
The size of the sample and of the unit cell is an in Figure \ref{azimuthal}.
}\label{air}
\end{figure}


\begin{thebibliography}{99}
\bibitem{kvak} We analyzed also LHM structures in with Real $\epsm$ large and negative
simultaneously with Im $\epsm$  large and positive (data not presented here). 
We found that the value of  Re $\epsm$  influences neither the position of 
the resonance  gap nor the absorption provided that Im $\epsm$ is large enough.
\bibitem{com} In our systems, the symmetry with respect to transformation
$x\to -x$ is broken by presence of the dielectric board with permittivity
$\eps=3.4$. This is a reason for non-zero (although very small)
values of the transmission $t(p\to s)$ and $t(s\to p)$ in the  ``turned'' SRR
array.
\end{thebibliography}
\end{document}